# Bias spectroscopy and simultaneous SET charge state detection of Si:P double dots

M. Mitic[1], K. D. Petersson[1], M. C. Cassidy[1], R. P. Starrett[1], E. Gauja[1], A. J. Ferguson[1],

C. Yang[2], D. N. Jamieson[2], R. G. Clark[1] and A. S. Dzurak[1`]

[1] Centre for Quantum Computer Technology, Schools of Electrical Engineering and Physics, The University of New South Wales, NSW 2052, Sydney Australia

[2] Centre for Quantum Computer Technology, School of Physics, University of Melbourne, VIC 3010, Australia

We report a detailed study of low-temperature (mK) transport properties of a silicon double-dot system fabricated by phosphorous ion implantation. The device under study consists of two phosphorous nanoscale islands doped to above the metal-insulator transition, separated from each other and the source and drain reservoirs by nominally undoped (intrinsic) silicon tunnel barriers. Metallic control gates, together with an Al-AlO$_x$ single-electron transistor, were positioned on the substrate surface, capacitively coupled to the buried dots. The individual double-dot charge states were probed using source-drain bias spectroscopy combined with non-invasive SET charge sensing. The system was measured in linear ($V_{SD} = 0$) and non-linear ($V_{SD} \neq 0$) regimes allowing calculations of the relevant capacitances. Simultaneous detection using both SET sensing and source-drain current measurements was demonstrated, providing a valuable combination for the analysis of the system. Evolution of the triple points with applied bias was observed using both charge and current sensing. Coulomb diamonds, showing the interplay between the Coulomb charging effects of the two dots, were measured using simultaneous detection and compared with numerical simulations.

## 1. Introduction

Double-dot systems[1] continue to attract significant interest due to the single-electron tunneling phenomena that governs their behavior and the versatility of their potential applications[1,2]. Semiconductor double *quantum* dots have been used extensively in the study of quantum states and their interactions[3,4] due to their potential for solid-state quantum computing[5,6]. Classical (many-electron) double-dots, on the other hand, have been proposed for use in single electron logic and memory circuits[2,7] and quantum cellular automata (QCA)[8,9].

In previous work, our group has developed a variety of single-electron devices based on nanoscale phosphorous-implanted dots in silicon[10,11,12]. Such devices are attractive due to their compatibility with scalable Si microfabrication technologies and the scope for miniaturization down to single phosphorous donors[13,14] leading to ultra dense, extremely low-power electronic devices. To date, silicon double-dots[10,11] and QCA cells[12], defined using phosphorous implantation, have been demonstrated, while there are ongoing efforts towards miniaturizing the dots[15] in order to access a regime where silicon-based quantum bits[16,17] may be demonstrated.

In this paper we present electrical measurements of a phosphorous ion-implanted silicon double-dot system, operated both in linear and non-linear regimes[1]. The double-dot charge states and their evolution with the applied source-drain DC bias ($V_{SD}$) were investigated using both SET charge sensing[18] and source-drain conductance measurements. Simultaneous detection using these two techniques was also achieved, allowing for a direct comparison between the two.

A schematic of the double-dot device is shown in Fig. 1a. It consists of two phosphorous-implanted nanoscale dots, tunnel coupled to each other and to two reservoirs labeled as source $s_A$ and drain $d_A$. The reservoirs allow characterisation of the device via direct current measurement and also enable the total number of electrons on the double-dot system to be adjusted. Control of the individual electron occupancy numbers $\{n_A, m_A\}$ within the double-dot system is provided by two surface gates, lebeled $V_L$ and $V_R$ in Fig.1a, while an Al-AlO$_x$ single electron transistor (SET), also positioned on the device surface, provides non-invasive double-dot charge-state sensing[19]. The devices also incorporate a second double-dot structure (upper section of Fig.1(b,c)) which was not used for the experiments reported here, but can be used for demonstration of quantum cellular automata[12].

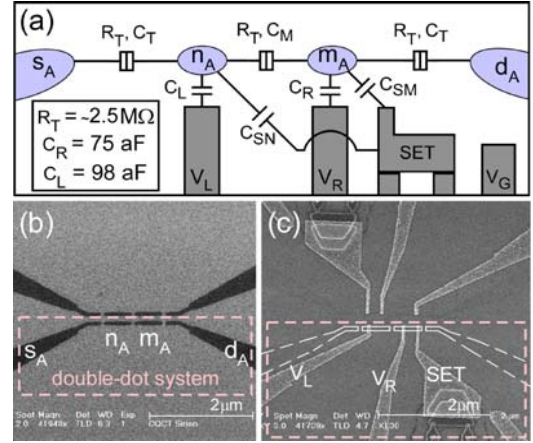

FIG. 1. a) Simplified circuit schematic of the double-dot device; b) SEM image of phosphorous-implanted double-dot with leads (n+ regions are dark in image); SEM image of the completed device. The buried double-dot with leads is marked using dashed lines. The upper half of the structures (outside the dashed box) were not used in these experiments.

## 2. Fabrication

The device was fabricated on a near-intrinsic (n < $10^{12}$ cm$^{-3}$) n-type silicon wafer of resistivity greater than 5 k$\Omega$cm. The n$^+$ ohmic contacts for the source and drain leads (labeled as $s_A$, $d_A$ in Fig. 1a) were first defined by phosphorous contact diffusion at 950C°, through a photo-lithographically defined oxide mask. This was followed by a deglazing HF etch, and a 200 nm drive-in oxidation, used to drive the diffused phosphorous deeper into the silicon substrate. The drive-in oxide was then removed by another HF etch, and a 5nm thick layer of high-quality SiO$_2$ was grown[20] by DCE–assisted (dichloroethylene) oxidation at 800 °C.

The dots and leads were formed by 14keV P$^+$ phosphorous ion implantation through an EBL patterned polymethyl–methacrylate (PMMA) resist mask. The dimensions of the dots were set to be 80x500nm, while the widths of the tunnel junctions, formed by the undoped near-intrinsic silicon substrate between the implanted regions, were set to 60nm (Fig 1b). The lateral straggle of

Mitic *et al.*

the phosphorous donors was found to be less than 20nm[13], hence the widths were significantly greater than the straggling. Typical tunneling junction resistance was found to be approximately 2.5 MΩ. The implantation produced a phosphorous doping density of ~$10^{19}$ $cm^{-3}$, an order of magnitude above the metal-insulator transition, with a mean depth of 15nm below the Si-SiO$_2$ interface[13]. To activate the implanted donors and to repair the damage in the substrate created by the implantation process, the devices were subjected to a rapid thermal anneal at 1000°C for 5 seconds.

Following the implantation and the activation of the donors, the control gates, labeled as $V_L$ and $V_R$ in Fig. 1a, were defined by EBL in a 60nm thick 950K PMMA resist and metalized on the device surface in 10nm of titanium followed by 20nm of gold. They were accurately positioned above the dots to ensure good capacitive coupling to them. The last step in the fabrication process was the formation of the SET. It was metalized by aluminum double-angle evaporation through a suspended EBL-patterned bilayer (PMMA/copolymer) resist mask[21], interrupted by a controlled oxidation step used to form the SET oxide tunneling barrier. The complete double-dot device with the metalized control gates and an SET is shown in Fig 1c.

Throughout the processing, an alignment accuracy of 25nm or better was ensured for each of the individual fabrication steps by using specially designed nanoscale registration markers[14]. The markers, 100nm x 100nm in size, were carefully engineered to survive the processing, including rapid thermal anneal temperatures of 1000°C. They were defined in PMMA resist by EBL, and metalized using 15nm of titanium followed by 65nm platinum.

### 3. Electrical measurements

All electrical measurements were carried out in a dilution refrigerator operated at a base temperature of ~50mK. Throughout the measurements, a magnetic field of 0.5 Tesla was applied in order to quench the superconductivity in aluminum metal. The SET was operated as a charge detector by applying compensating voltage to the bias gate (labeled as $V_G$ in Fig. 1a) in order to counter the electrostatic potential changes induced on the island of the SET by the varying control gate voltages.

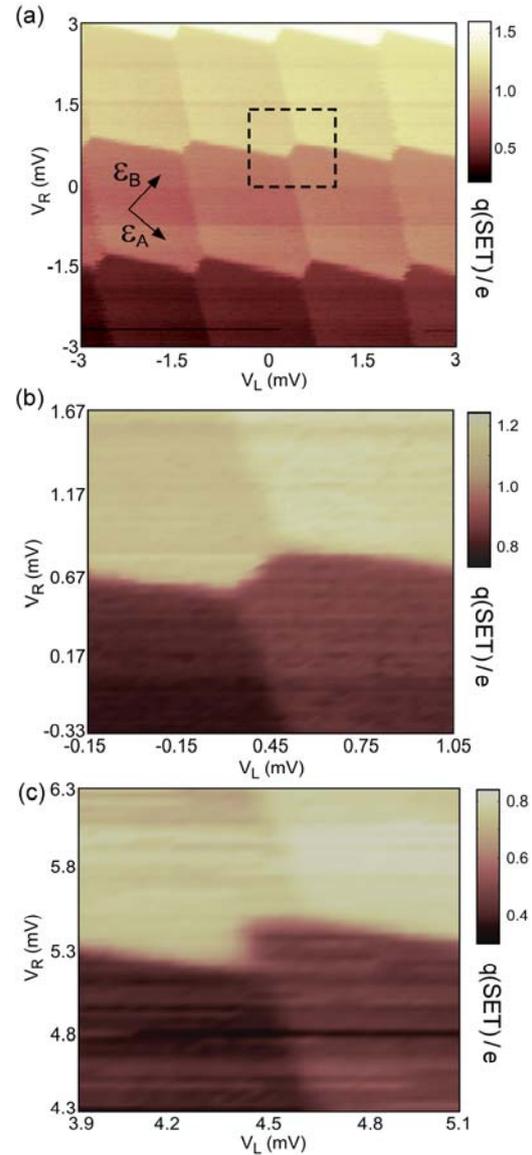

FIG. 2. Normalized charge q/e induced on the island of the SET as a function of the control gate voltages $V_L$ and $V_R$. A plane was subtracted from the raw data to remove direct gate-SET coupling. (a) Data for $V_{SD} = 0$, showing characteristic "honeycomb" charge-state diagram; (b) zoom-in showing four arbitrary charge states and the border separating them at $V_{SD} = 0$; (c) similar plot for $V_{SD} = 100\mu V$;

To characterize the device, the double dot electron occupancy states, measured in a linear transport regime ($V_{SD} = 0$), were first determined by sweeping control gate voltages $V_L$ and $V_R$ and monitoring the state of the system using the SET. The normalized charge induced on the island of the SET is shown in Fig. 2a. The honeycomb structure observed is typical of double-dot systems[1]. The individual states are labeled by the relative electron occupancy numbers of the two dots ($n_A$, $m_A$). The SET is evidently more sensitive to changes in $m_A$ than changes in $n_A$ as its capacitive coupling to the right dot is greater due to the device geometry. The data obtained showed reasonable stability over time. Using the sizes of the individual state cells, the gate-to-dot capacitances were calculated[1] to be $C_L$ = 98 aF and $C_R$ = 75 aF.

The state space of the double-dot was also measured in a non-linear transport regime, by applying a DC bias voltage across the source and drain reservoirs using a battery power supply. Typical SET charge sensor responses for the two cases of no bias and 100 μV DC bias applied across the source and drain electrodes are shown in Fig. 2b and 2c respectively. The border between the states ($n_A$, $m_A$) and ($n_A$+1, $m_A$-1), in Fig. 2c, has been modified by the applied DC bias, as the triple points extend into triangles. From these extensions, the total capacitances of the left and right dot were calculated[1] to be 190aF and 180aF respectively (corresponding to individual dot charging energies of approximately 400 μV), while the coupling capacitance between the two dots $C_m$ was found to be ~30aF using the linear transport data in Fig 2a.

In order to better observe the triple points and their evolution with the applied DC bias, the source-drain conductance of the double dot was measured using a standard lock-in technique. The measurement was done using a 20 μV AC voltage signal applied at 78Hz, combined with the DC offset voltage of variable value. The resulting plots of the source-drain current through the double dot as a function of control gate polarizations $\varepsilon_A$ and $\varepsilon_B$, for different values of DC bias applied, are shown in Fig. 3. The control gate polarizations are shown in Figure 2a and are defined as: $\varepsilon_A = \cos(\theta)V_L - \sin(\theta)V_R$, $\varepsilon_B = \cos(\theta)V_R - \sin(\theta)V_L$, where $\theta \approx \pi/4$.

As predicted by theory[1], the size of the triple point triangles shows approximate linear dependence on source drain bias applied across the double-dot. The absolute size of the triangles holds the information about the total capacitance of each dot, but due to the unavoidable small

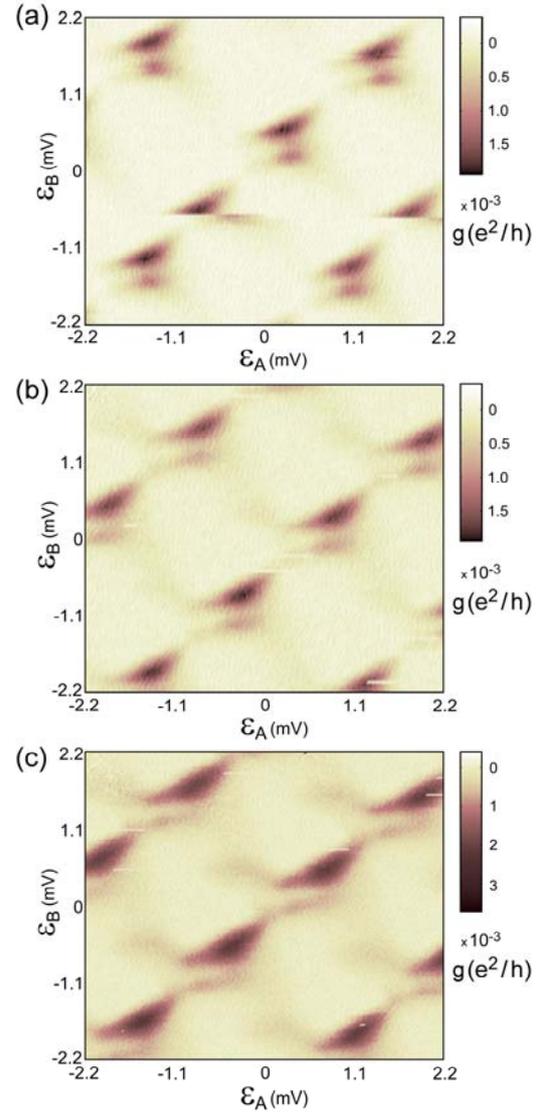

FIG. 3. Normalized source-drain conductance g of the double-dot device as a function of gate polarizations $\varepsilon_A$ and $\varepsilon_B$, measured at a variety of source-drain bias voltages. (a) $V_{SD}$ = 10 μV, (b) $V_{SD}$ = 60 μV, (c) $V_{SD}$ = 150 μV.

source-drain DC voltage offset present during the measurements, the capacitances were calculated more accurately from the relative size increases. The capacitances of the left and right dot were found to be 250 and 50 aF respectively, within the order of magnitude of the SET measurements previously done. In comparison to the SET charge sensing, source–drain current measurements have been considerably more noisy, due to the fact that the current measured through the double-dot at a finite DC source-

drain bias is more susceptible to electron tunneling through the random traps and impurities found in the substrate, around the non-ideal edges of the implanted dots and in the Si-$SiO_2$ interface. The fact that the upper triple point triangle appears larger in size is possibly due to the nonuniform internal structure of the individual dots.

To confirm that the regions of high current, shown in Fig. 3, truly correspond to the triple points shown in Fig. 2, the simultaneous detection of the electron occupancy states of the double dot was attempted, using the SET charge detection and the source-drain conductance measurement. Figure 4 below shows the resulting induced charge on the island of the SET and the simultaneously measured conductance, for the case of 60 µV DC bias applied across the double dot. The matching between the two is good, with the SET signal picking up more noise than in previous cases (Fig. 2), due to both DC and AC signals being applied across the double-dot.

The technique of simultaneously measuring the current through the double-dot and the response of the capacitively coupled SET was also used to directly determine the charging energies of the dots. This was done by sweeping the control gate voltage on the left dot for different values of DC bias applied across the source and the drain electrodes of the double-dot. Regular Coulomb diamond structures were observed both in SET charge sensor response (Fig. 5a) and the simultaneously measured source-drain conductance (Fig. 5b). The data revealed an interaction of Coulomb charging effects of the two dots, as the left control gate was also weakly capacitively coupled to the right dot. This is evident from the variation in the size of the blockaded region, (see Fig. 5) as the right dot is driven from the non-blockaded to the fully blockaded state by the control gate voltage. The measurements revealed a typical dot charging energy of approximately 200 µV, lower then the values obtained previously but still within the correct order of magnitude.

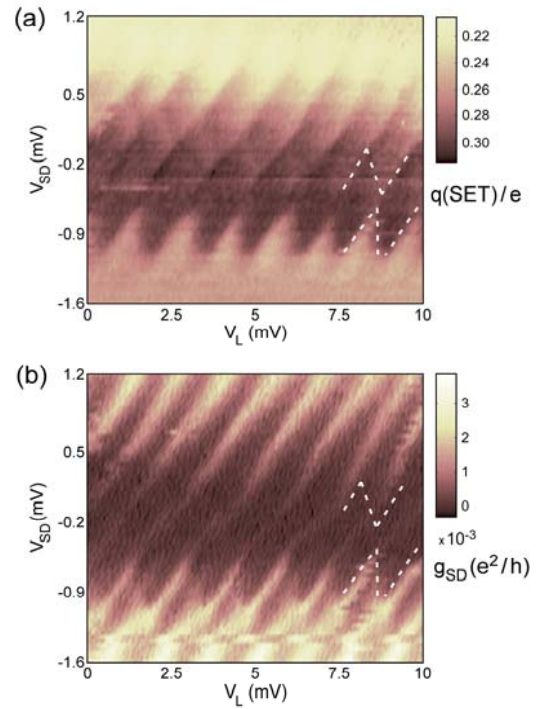

FIG. 5. Simultaneous measurements of double-dot Coulomb charging diamonds. (a) Normalized charge q induced on the island of the SET and (b) the source-drain conductance g, as functions of left control gate voltage and the source-drain bias applied.

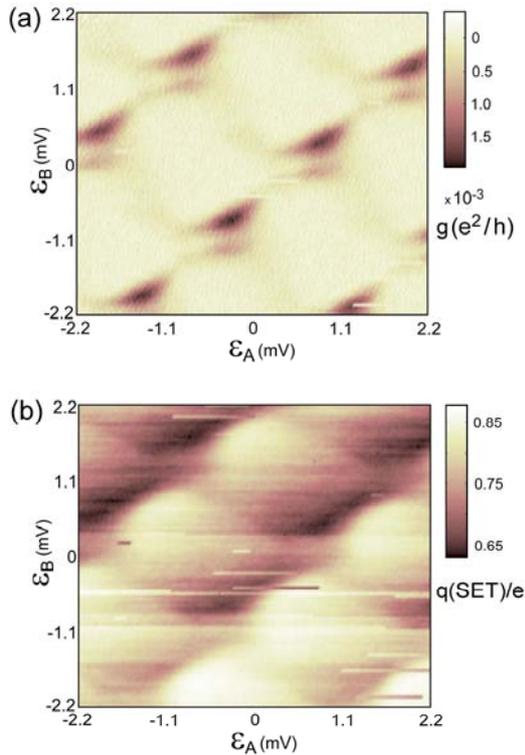

FIG. 4. Simultaneous measurements of the double-dot device at $V_{SD}$=60µV. (a) The source-drain conductance g, and (b) the normalized charge q induced on the island of the SET, as functions of gate polarizations $\varepsilon_A$ and $\varepsilon_B$.

SIMON software was used to model the experiment (the SETs and other circuitry were not included in the model). The resulting stability plots are shown in figure 6 below.

Reasonable consistency with the experimental data was achieved. The typical dot charging energy was found to be approximately 500μV, slightly higher then the measured values, which was to be expected due to the oversimplification of the model used.

## 4. Conclusions

In summary, the transport properties of Si:P double dot have been studied by using SET charge sensing and source-drain conductance measurements. The double dot has been operated both in linear and non-linear regimes and the relevant capacitances have been calculated from the data. The source-drain conductance measurements have proven to be significantly more noisy due electron tunneling through random traps and impurities. Simultaneous measurements using the two detection techniques have also been demonstrated showing good matching between the two. This has been used to observe the combined Coulomb charging effects of the two dots and to measure and confirm their individual charging energies. The double-dot system was aslo modelled using SIMON and the data was consistent with the experiment.

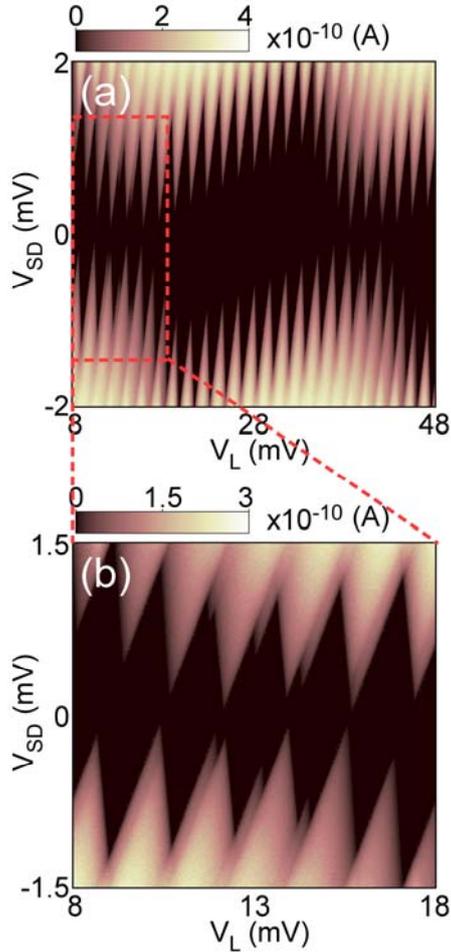

FIG. 6. SIMON model of charge oscillations in the double-dot device system as a function of the source-drain bias and the left control gate voltage. The data reveals a superposition of Coulomb charging effects of the two dots with a typical dot charging energy of approximately 500 μV.

### Acknowledgments

The authors would like to thank A. Fuhrer and A. R. Hamilton for helpful discussions. We would also like to thank G. Tamanyan for assistance with the ion implantations. This work was supported by the Australian Research Council, the Australian Government, the US Advanced Research and Development Activity, National Security Agency and Army Research Office under contract DAAD19-01-1-0653.